\pdfoutput=1
\PassOptionsToPackage{numbers}{natbib}
\documentclass[12pt]{article}

\usepackage[preprint]{neurips_2025}
\usepackage{times}
\usepackage[T1]{fontenc}
\usepackage[utf8]{inputenc}
\usepackage{array}
\usepackage{multirow}
\usepackage{booktabs}
\usepackage{amsmath}
\usepackage{amssymb}
\usepackage{pifont}

\usepackage[protrusion=false, expansion=false]{microtype}

\usepackage{hyperref}
\usepackage{url}
\usepackage{booktabs}
\usepackage{amsfonts}
\usepackage{nicefrac}
\usepackage[table]{xcolor}
\usepackage{adjustbox}
\usepackage{rotating}
\usepackage{parskip}
\usepackage{enumitem}
\usepackage{setspace}
\usepackage{titlesec}
\usepackage{etoolbox}

\titlespacing*{\section}{0pt}{0.3em plus 0.1em minus 0.1em}{0.3em}
\titlespacing*{\subsection}{0pt}{0.3em plus 0.1em minus 0.1em}{0.3em}
\titlespacing*{\subsubsection}{0pt}{0.8em}{0.5em}

\titleformat{\section}
  {\normalfont\fontsize{14pt}{17pt}\bfseries}
  {\thesection}{1em}{}

\titleformat{\subsection}
  {\normalfont\fontsize{12pt}{15pt}\bfseries}
  {\thesubsection}{1em}{}

\titleformat{\subsubsection}
  {\normalfont\fontsize{11pt}{14pt}\bfseries}
  {\thesubsubsection}{1em}{}

\title{PJB: A Reasoning-Aware Benchmark for Person-Job Retrieval}
\usepackage{graphicx}
\usepackage{tikz}
\usetikzlibrary{positioning, arrows.meta, calc}

\usepackage{listings}
\usepackage{tcolorbox}
\tcbuselibrary{listings,breakable}
\newtcolorbox{promptblock}{
  colback=gray!8,
  colframe=gray!50,
  boxrule=0.4pt,
  arc=2pt,
  left=6pt, right=6pt, top=4pt, bottom=4pt,
  breakable,
}

\author{%
  \textbf{Guangzhi Wang} \quad \textbf{Xiaohui Yang} \quad \textbf{Kai Li} \quad \textbf{Jiawen He} \quad \textbf{Kai Yang} \quad \textbf{Ruixuan Zhang} \\[0.5em]
  \textbf{Zhi Liu} \\[0.5em]
  CareerInternational Research Team \\[0.5em]
}

\begin{document}
\maketitle

\begin{abstract}
As retrieval models converge on generic benchmarks, the pressing question is no longer ``who scores higher'' but rather ``where do systems fail, and why?'' Person-job matching is a domain that urgently demands such diagnostic capability---it requires systems not only to verify explicit constraints but also to perform skill-transfer inference and job-competency reasoning, yet existing benchmarks provide no systematic diagnostic support for this task. We introduce PJB (Person-Job Benchmark), a reasoning-aware retrieval evaluation dataset that uses complete job descriptions as queries and complete resumes as documents, defines relevance through job-competency judgment, is grounded in real-world recruitment data spanning six industry domains and nearly 200{,}000 resumes, and upgrades evaluation from ``who scores higher'' to ``where do systems differ, and why'' through domain-family and reasoning-type diagnostic labels. Diagnostic experiments using dense retrieval reveal that performance heterogeneity across industry domains far exceeds the gains from module upgrades for the same model, indicating that aggregate scores alone can severely mislead optimization decisions. At the module level, reranking yields stable improvements while query understanding not only fails to help but actually degrades overall performance when combined with reranking---the two modules face fundamentally different improvement bottlenecks. The value of PJB lies not in yet another leaderboard of average scores, but in providing recruitment retrieval systems with a capability map that pinpoints where to invest.
\end{abstract}

\section{Introduction}

\begin{sloppypar}
The competitive focus of AI is shifting from pretraining scale and general capability demonstrations toward utility verification and system deployment for real-world tasks \citep{yao2025secondhalf}. For evaluation, this shift means that benchmarks must go beyond answering ``how strong is the model on average'' and further address ``whether the model is truly useful in complex, high-stakes, inference-demanding real-world scenarios.'' The development of retrieval benchmarks clearly reflects this evolution: BEIR \cite{thakur2021beir} and MTEB \citep{muennighoff-etal-2023-mteb} expanded evaluation from single datasets to cross-domain, cross-task zero-shot generalization; BRIGHT \citep{su2024bright} further focused on reasoning-intensive retrieval, probing whether systems can handle complex queries that cannot be resolved through lexical matching or shallow semantic similarity alone. However, these benchmarks remain largely confined to general or semi-general retrieval scenarios and have yet to systematically address recruitment---a real-world task characterized by long documents, multiple constraints, and job-competency-driven relevance judgments.
\end{sloppypar}

\begin{sloppypar}
Person-job matching is precisely such a compound retrieval problem: on the surface, it retrieves suitable candidates from job descriptions, but in reality it simultaneously involves two types of reasoning demands. Parallel reasoning requires systems to independently verify explicit constraints---location, education, years of experience, salary range, and job keywords---and synthesize the results, a class of multi-constraint retrieval problems that has received increasing attention in recent compound retrieval research \citep{killingback2025crumb}. Serial reasoning, by contrast, requires systems to perform multi-hop semantic abstraction around job competency---for example, mapping job responsibilities to implicit skill requirements, interpreting cross-industry experience as transferable capabilities, and aggregating candidate evidence across multiple resume sections---bearing structural similarity to the cross-document evidence chain reasoning studied in multi-hop retrieval \citep{tang2024multihoprag}. Because both types of reasoning coexist, person-job matching can be reduced to neither structured filtering nor standard text similarity ranking; nor can a single aggregate metric adequately characterize system utility on real business queries. To evaluate such systems, a benchmark must go beyond overall ranking quality and explain where systems fail---across which domain families, reasoning types, and low-gain queries.
\end{sloppypar}

\begin{sloppypar}
To this end, we propose PJB (Person-Job Benchmark), a reasoning-aware benchmark that formalizes the matching of complete job descriptions and complete resumes as a reproducible, diagnostic offline retrieval evaluation dataset, designated PJB v1.0. PJB v1.0 is constructed from real recruitment data and, like mainstream retrieval benchmarks, adopts a fixed query set, fixed document corpus, and fixed relevance judgments in an offline evaluation setting. Through domain-family and reasoning-type diagnostic labels, the benchmark goes beyond reporting aggregate scores to localize system capability structures. Specifically, the contributions of this paper are as follows:

\begin{enumerate}[leftmargin=2em,itemsep=2pt,parsep=0pt]
  \item We formalize person-job matching as a job-competency-driven retrieval task and construct an evaluation dataset comprising nearly 300 queries, nearly 200{,}000 resumes, and over 2{,}000 positive relevance judgments, along with domain-family and reasoning-type label systems that support diagnostic analysis.
  \item Using dense retrieval as the entry point, we conduct unified evaluation across different model versions and module combinations, revealing pronounced domain heterogeneity and reasoning heterogeneity in person-job retrieval, with the reranking module being the most stable source of improvement.
  \item On the data construction and usage boundary, we address the bias \citep{wilson2024gender,vladimirova2024fairjob} and privacy risks \citep{yamashita2024openresume} specific to recruitment scenarios through de-identification and compliance measures, described in the method section.
\end{enumerate}

\noindent These results demonstrate that PJB can serve as a more business-relevant unified baseline for comparing, diagnosing, and subsequently optimizing recruitment retrieval systems.
\end{sloppypar}

\section{Related Work}

\begin{sloppypar}
From the perspective of retrieval evaluation evolution, PJB inherits a lineage from fixed-collection scoring toward cross-domain and diagnostic benchmarks. Cranfield established the offline evaluation paradigm of fixed document collections, fixed query sets, and fixed relevance judgments \citep{cleverdon1966factors}, and TREC extended this paradigm to large-scale pooling-based evaluation \citep{voorhees2005trec}. Subsequent work addressed the incomplete judgment problem inherent in pooling by proposing more robust methods such as bpref and sampling-based estimation, aiming to reduce the bias from treating unjudged documents as irrelevant \citep{buckley2004retrieval,yilmaz2008simple}. With the advent of representation learning and universal embedding models, benchmark focus shifted further from aggregate scores on a single collection toward cross-dataset, cross-task, and zero-shot generalization, with BEIR and MTEB representing this heterogeneous evaluation turn \citep{thakur2021beir,muennighoff-etal-2023-mteb}; BRIGHT then pushed this trajectory toward reasoning-intensive retrieval, emphasizing that benchmarks should distinguish not only ``whether something can be retrieved'' but also ``whether the system can handle queries requiring inference and evidence organization'' \citep{su2024bright}. Concurrently, LLM-as-a-Judge has begun to alleviate the cost bottleneck of large-scale relevance annotation, though it introduces its own calibration and bias risks \citep{rahmani2025judging}. Overall, retrieval evaluation has increasingly prioritized cross-domain generalization and complex reasoning capabilities, yet rarely anchors the evaluation target in real-world person-job matching---a long-document, multi-constraint, competency-judgment scenario.
\end{sloppypar}

\begin{sloppypar}
From the application perspective, person-job matching research itself has evolved from structured feature alignment to semantic matching, and further to knowledge-enhanced and LLM-assisted systems. Early systems relied predominantly on explicit fields---education, years of experience, skill keywords, and job titles---for rule-based matching or ranking, which handled enumerable constraints effectively but struggled with cross-industry experience transfer, implicit skill mapping, and long-text responsibility understanding. As the need for modeling job descriptions and resume text grew, PJFNN \citep{zhu2019pjf} and APJFNN \citep{qin2020apjfnn} formulated person-job matching as joint representation learning and interaction-aware matching between job text and resume text, enabling models to learn finer-grained competency alignment than manual features. Subsequent work further incorporated skill--occupation graph context and LLM distillation, attempting to bridge the implicit semantic gap between job requirements and candidate experience, embedding person-job matching capabilities into broader HR NLP pipelines \citep{pezeshkpour2023rjdb}. However, this field remains largely organized around paired scoring or ranking optimization on proprietary data, with evaluation protocols, annotation standards, and error analysis dimensions varying across tasks and organizations, making it difficult to compare method improvements on a unified benchmark or to disentangle ``overall ranking quality'' from ``failure modes in specific domains or reasoning types.'' PJB is positioned precisely at the intersection of these two lines: combining mature retrieval evaluation paradigms with real person-job matching scenarios to construct a benchmark that is both reproducible and capable of supporting diagnostic analysis.
\end{sloppypar}

\section{Method}
This section introduces PJB's task definition and data composition, diagnostic label taxonomy, construction pipeline for relevance judgments, and the design rationale of the evaluation protocol. PJB adopts the offline evaluation paradigm of fixed query sets, fixed document corpora, and fixed relevance judgments, specialized for complete job description and complete resume matching in Chinese recruitment scenarios, enabling the benchmark to go beyond aggregate scores and explain system capability differences across business slices.

\subsection{Overview}\label{sec:overview}

PJB defines person-job matching as a query--document retrieval task: the query is a complete job description, the document is a complete resume, and the system objective is to return a relevance-ranked candidate list from a fixed resume corpus for each JD. Relevance here is defined not as ``whether an offer would certainly be extended'' or ``whether an interview would certainly follow,'' but rather as a more stable job-competency judgment---whether the candidate demonstrates sufficient evidence of competency and experience to qualify for the position. This dataset is designated PJB v1.0. The current v1.0 comprises nearly 300 queries, nearly 200{,}000 resumes, and over 2{,}000 binary positive relevance judgments; all queries have at least one positive, with an average of approximately eight positives per query, making it a typical multi-positive, sparse-label retrieval benchmark. Table~\ref{tab:data-structure} provides the key scale statistics.

\begin{table}[ht]
  \centering
  \caption{Scale statistics of PJB v1.0.}
  \label{tab:data-structure}
  \small
  \begin{tabular}{ll}
    \toprule
    Statistic & Value \\
    \midrule
    Number of Queries & 297 \\
    Document Corpus Size & 197{,}674 \\
    Total Positive Judgments & 2{,}242 \\
    Avg. Positives per Query & 7.55 \\
    Document/Query Ratio & 665.57 \\
    Positive Density & 0.0038\,\% \\
    \bottomrule
  \end{tabular}
\end{table}

\begin{sloppypar}
As illustrated in Figure~\ref{fig:build-flow}, PJB's construction involves three stages. First, the query set and document corpus are sourced from internal search logs after 2025-01-01, with queries consisting of complete JDs and documents consisting of de-identified complete CVs. Both are composite objects containing structured slots and long-text fields: queries include job category, location, education, experience, salary range, and responsibility descriptions; documents include education history, work history, project experience, and job preferences. Second, the candidate generation stage uses BM25 and multiple dense retrieval pipelines to select high-potential query--document pairs, after which a two-stage LLM-as-a-Judge pipeline---with doubao-1.5 performing initial filtering and kimi-2.5 providing binary job-competency judgments \citep{rahmani2025judging}---is applied, supplemented by manual spot-checking on approximately 20\% of queries. Finally, the fixed query set, fixed document corpus, and fixed relevance judgments together constitute the offline evaluation dataset; the released \texttt{qrels.tsv} retains only $\text{rel}=1$ positive judgments, and missing pairs cannot be distinguished as judged negatives or candidates that never entered the pool. The domain families and reasoning types mentioned subsequently are auxiliary diagnostic labels used only for sliced analysis; see Section~\ref{sec:taxonomy}.
\end{sloppypar}

% --- Paper color definitions ---
\definecolor{paperBlue}{RGB}{52, 116, 172}
\definecolor{paperGreen}{RGB}{79, 173, 151}
\definecolor{paperOrange}{RGB}{231, 151, 92}

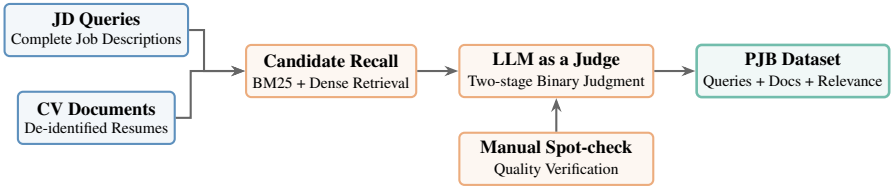
\begin{figure}[ht]
  \centering
  \resizebox{0.844\columnwidth}{!}{%
  \begin{tikzpicture}[
    node distance=0.45cm and 0.55cm,
    base/.style={
        draw, thick, rounded corners=2pt,
        align=center,
        font=\rmfamily\scriptsize,
        inner sep=2.5pt,
        minimum width=1.9cm,
        minimum height=0.7cm,
        fill=white
    },
    srcNode/.style={base, fill=paperBlue!6, draw=paperBlue!80},
    procNode/.style={base, fill=paperOrange!8, draw=paperOrange!80},
    outNode/.style={base, fill=paperGreen!8, draw=paperGreen!80, line width=1pt},
    arr/.style={-{Stealth[scale=0.8]}, thick, draw=black!60}
  ]

    % --- Data sources ---
    \node[srcNode] (jd) {\textbf{JD Queries}\\[-1pt]\tiny Complete Job Descriptions};
    \node[srcNode, below=of jd] (cv) {\textbf{CV Documents}\\[-1pt]\tiny De-identified Resumes};

    % --- Candidate recall ---
    \node[procNode, right=of jd, xshift=0.2cm, yshift=-0.55cm] (recall) {
        \textbf{Candidate Recall} \\
        \tiny BM25 + Dense Retrieval
    };

    % --- LLM Judge ---
    \node[procNode, right=of recall, minimum width=2.6cm] (judge) {
        \textbf{LLM as a Judge} \\
        \tiny Two-stage Binary Judgment
    };

    % --- Manual QA ---
    \node[procNode, below=of judge, minimum width=2.6cm] (qa) {
        \textbf{Manual Spot-check} \\
        \tiny Quality Verification
    };

    % --- Final output ---
    \node[outNode, right=of judge] (out) {
        \textbf{PJB Dataset} \\
        \tiny Queries + Docs + Relevance
    };

    % --- Edges ---
    \draw[arr] (jd.east) -- ++(0.2,0) |- (recall.west);
    \draw[arr] (cv.east) -- ++(0.2,0) |- (recall.west);
    \draw[arr] (recall) -- (judge);
    \draw[arr] (judge) -- (out);
    \draw[arr] (qa) -- (judge);

  \end{tikzpicture}%
  }
  \caption{Construction pipeline of PJB.}
  \label{fig:build-flow}
\end{figure}

\subsection{Taxonomy}
\label{sec:taxonomy}

Beyond relevance judgments, PJB augments queries with diagnostic labels, upgrading the benchmark from aggregate evaluation to diagnostic evaluation. These labels provide stable business-oriented slices on top of the same set of relevance judgments, enabling aggregation and interpretation of system performance. Two types of query-side diagnostic labels are currently used: domain families and reasoning types.

\subsubsection{Domain Taxonomy}

The raw query-side data contains over 30 fine-grained job categories. Using them directly for grouped evaluation would result in most buckets having insufficient sample sizes and unstable statistics. We therefore aggregate them into six domain families based on the similarity of their core matching competencies. The aggregation principle is: if two job categories rely on the same competency dimensions when screening candidates, they are grouped into the same domain family. Specifically, Technical R\&D aggregates software development, algorithms, testing, and operations roles that use technology stacks and engineering experience as core criteria; Product \& Operations aggregates product managers and various operations roles centered on business understanding and user growth; HR/Admin/Finance aggregates functional roles driven by institutional and procedural requirements; Sales \& Market Support aggregates commercial roles centered on client communication and industry resources; Mechanical/Hardware aggregates hardware engineering roles requiring domain-specific expertise; and Project Management covers roles centered on cross-team coordination. Since the matching logic within each domain family is internally consistent while inter-family differences are pronounced, this aggregation level ensures sufficient statistical samples per bucket while preserving business-level interpretability. Figure~\ref{fig:query-family-pie} shows the distribution of queries across the six domain families.

% --- Professional color palette ---
\definecolor{color1}{RGB}{52, 116, 172}   % Tech Blue (Tech. R&D)
\definecolor{color2}{RGB}{79, 173, 151}   % Forest Green (Product)
\definecolor{color3}{RGB}{231, 151, 92}   % Warm Orange (HR)
\definecolor{color4}{RGB}{204, 82, 72}    % Brick Red (Sales)
\definecolor{color5}{RGB}{145, 125, 182}  % Lilac Purple (Mech)
\definecolor{color6}{RGB}{119, 186, 230}  % Light Sky Blue (Proj)

\begin{figure}[ht]
  \centering
  \begin{tikzpicture}[scale=1.0]

    \def\radius{1.6}
    \def\innerradius{0}

    % Starting angle definitions
    \def\angA{0}
    \def\angB{144.24}
    \def\angC{207.27}
    \def\angD{256.97}
    \def\angE{306.67}
    \def\angF{350.30}
    \def\angEnd{360}

    % --- Draw sectors ---
    \begin{scope}[line join=round, line cap=round]
        \fill[color1, draw=white, line width=0.8pt] (0,0) -- (\angA:\radius) arc (\angA:\angB:\radius) -- cycle;
        \fill[color2, draw=white, line width=0.8pt] (0,0) -- (\angB:\radius) arc (\angB:\angC:\radius) -- cycle;
        \fill[color3, draw=white, line width=0.8pt] (0,0) -- (\angC:\radius) arc (\angC:\angD:\radius) -- cycle;
        \fill[color4, draw=white, line width=0.8pt] (0,0) -- (\angD:\radius) arc (\angD:\angE:\radius) -- cycle;
        \fill[color5, draw=white, line width=0.8pt] (0,0) -- (\angE:\radius) arc (\angE:\angF:\radius) -- cycle;
        \fill[color6, draw=white, line width=0.8pt] (0,0) -- (\angF:\radius) arc (\angF:\angEnd:\radius) -- cycle;
    \end{scope}

    % --- Legend ---
    \begin{scope}[shift={(2.2, 1.5)}]
        \foreach \c/\t/\v [count=\i] in {
            color1/{Tech. R\&D}/119,
            color2/{Product \& Ops}/52,
            color3/{HR/Admin/Fin.}/41,
            color4/{Sales \& Support}/41,
            color5/{Mech./Hardware}/36,
            color6/{Proj. Mgmt}/8
        } {
          \coordinate (row) at (0, -\i*0.42);
          \fill[\c] (0.8, -\i*0.42 - 0.1) rectangle ++(0.25, 0.18);
          \node[anchor=west, font=\scriptsize, text=black!90] at (1.0, -\i*0.42) {\t};
          \node[anchor=east, font=\scriptsize, text=black!90] at (4.2, -\i*0.42) {\v};
        }
    \end{scope}

  \end{tikzpicture}
  \caption{Distribution of queries across the six domain families in PJB.}
  \label{fig:query-family-pie}
\end{figure}
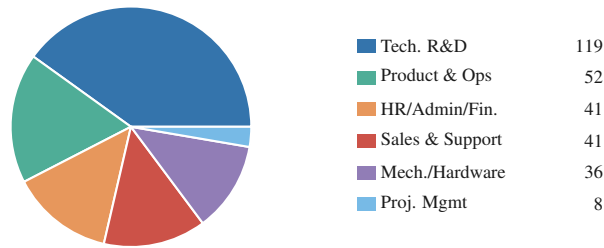

\subsubsection{Reasoning Taxonomy}

Beyond domain slicing, PJB also assigns reasoning-type labels to queries through heuristic rules, characterizing a system's ability to handle queries of varying complexity. This design shares motivation with reasoning-intensive retrieval benchmarks such as BRIGHT: retrieval difficulty depends not only on semantic similarity but also on the system's ability to organize evidence and perform inference \citep{su2024bright}.

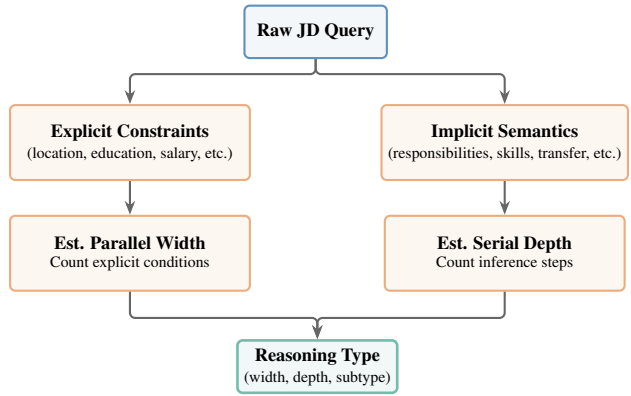
\begin{figure}[ht]
  \centering
  \resizebox{0.590\columnwidth}{!}{%
  \begin{tikzpicture}[
    node distance=0.45cm and 0.55cm,
    base/.style={
        draw, thick, rounded corners=2pt,
        align=center,
        font=\rmfamily\scriptsize,
        inner sep=2.5pt,
        minimum width=1.9cm,
        minimum height=0.7cm,
        fill=white
    },
    nodeIn/.style={base, fill=paperBlue!6, draw=paperBlue!80},
    nodeProc/.style={base, fill=paperOrange!8, draw=paperOrange!80},
    nodeOut/.style={base, fill=paperGreen!8, draw=paperGreen!80, line width=1pt},
    arr/.style={-{Stealth[scale=0.8]}, thick, draw=black!60, rounded corners=2pt}
  ]

    \node[nodeIn] (A) {\textbf{Raw JD Query}};

    \node[nodeProc, minimum width=3.2cm, minimum height=1.0cm, below left=0.6cm and -0.1cm of A] (B) {
        \textbf{Explicit Constraints} \\
        \tiny (location, education, salary, etc.)
    };

    \node[nodeProc, minimum width=3.2cm, minimum height=1.0cm, below right=0.6cm and -0.1cm of A] (C) {
        \textbf{Implicit Semantics} \\
        \tiny (responsibilities, skills, transfer, etc.)
    };

    \node[nodeProc, minimum width=3.2cm, minimum height=1.0cm, below=of B] (D) {\textbf{Est. Parallel Width}\\[-1pt]\tiny Count explicit conditions};
    \node[nodeProc, minimum width=3.2cm, minimum height=1.0cm, below=of C] (E) {\textbf{Est. Serial Depth}\\[-1pt]\tiny Count inference steps};

    \node[nodeOut] (F) at ($(D.south)!0.5!(E.south) + (0,-1.0)$) {
        \textbf{Reasoning Type} \\
        \tiny (width, depth, subtype)
    };

    \draw[arr] (A.south) -- ++(0,-0.2) -| (B.north);
    \draw[arr] (A.south) -- ++(0,-0.2) -| (C.north);
    \draw[arr] (B) -- (D);
    \draw[arr] (C) -- (E);
    \draw[arr] (D.south) -- ++(0,-0.35) -| (F.north);
    \draw[arr] (E.south) -- ++(0,-0.35) -| (F.north);

  \end{tikzpicture}%
  }
  \caption{Heuristic estimation pipeline for PJB reasoning types.}
  \label{fig:reasoning-type-estimation}
\end{figure}

Specifically, PJB extracts two numerical dimensions from each query (as shown in Figure~\ref{fig:reasoning-type-estimation}). Parallel width counts the number of independently verifiable explicit constraints in the query---location, education, salary, years of experience---which are mutually independent and can be checked one by one before intersection. Serial depth estimates the number of additional semantic normalization and multi-step reasoning steps the system must perform to assess job competency---for example, mapping responsibility descriptions to implicit skill requirements, or inferring transferable capabilities from cross-industry experience. Based on the combination of these two dimensions, queries are classified into three types: queries with parallel width $\geq 3$ and serial depth $= 0$ are classified as \emph{parallel-only}, requiring only joint filtering of explicit constraints; queries with both parallel width and serial depth $\geq 1$ are classified as \emph{hybrid-balanced}, demanding both explicit filtering and semantic inference; queries with serial depth $\geq 2$ and low parallel width are classified as \emph{serial-dominant}, requiring the system to perform substantial cross-field inference to identify positives.

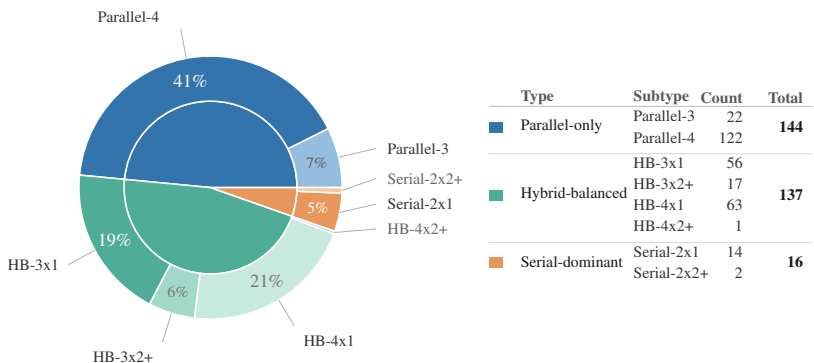
\begin{figure}[ht]
  \centering
  \begin{tikzpicture}
    % === Colors ===
    \definecolor{cPar}{RGB}{52, 116, 172}
    \definecolor{cParL}{RGB}{149, 189, 224}
    \definecolor{cHyb}{RGB}{79, 173, 151}
    \definecolor{cHybL}{RGB}{162, 216, 199}
    \definecolor{cHybLL}{RGB}{199, 233, 222}
    \definecolor{cSer}{RGB}{231, 151, 92}
    \definecolor{cSerL}{RGB}{245, 200, 160}

    % === Radii ===
    \def\Rm{1.15}
    \def\Ro{1.75}

    % === Solid sector macro ===
    \newcommand{\piesec}[4]{%
      \fill[#4, draw=white, line width=0.6pt]
        (0,0) -- (#1:#3) arc (#1:#2:#3) -- cycle;
    }
    % === Ring sector macro ===
    \newcommand{\ringsec}[5]{%
      \fill[#5, draw=white, line width=0.6pt]
        (#1:#3) arc (#1:#2:#3) -- (#2:#4) arc (#2:#1:#4) -- cycle;
    }

    % ============ Shift pie chart left ============
    \begin{scope}[shift={(-2.2, 0)}]
    % ============ Inner ring: three major reasoning types ============
    \piesec{0}{174.55}{\Rm}{cPar}
    \piesec{174.55}{340.61}{\Rm}{cHyb}
    \piesec{340.61}{360}{\Rm}{cSer}

    % ============ Outer ring: eight reasoning subtypes ============
    \ringsec{0}{26.67}{\Rm}{\Ro}{cParL}        % Parallel-3
    \ringsec{26.67}{174.55}{\Rm}{\Ro}{cPar}     % Parallel-4
    \ringsec{174.55}{242.42}{\Rm}{\Ro}{cHyb}    % HB-3x1
    \ringsec{242.42}{263.03}{\Rm}{\Ro}{cHybL}   % HB-3x2+
    \ringsec{263.03}{339.39}{\Rm}{\Ro}{cHybLL}  % HB-4x1
    \ringsec{339.39}{340.61}{\Rm}{\Ro}{cHybL}   % HB-4x2+
    \ringsec{340.61}{357.58}{\Rm}{\Ro}{cSer}    % Serial-2x1
    \ringsec{357.58}{360}{\Rm}{\Ro}{cSerL}      % Serial-2x2+

    % Outer ring percentages
    \pgfmathsetmacro{\Rp}{(\Rm+\Ro)/2}
    \node[font=\tiny, text=black!65] at (13.34:\Rp) {7\%};
    \node[font=\scriptsize, text=white] at (100.61:\Rp) {41\%};
    \node[font=\scriptsize, text=white] at (208.49:\Rp) {19\%};
    \node[font=\tiny, text=black!55] at (252.73:\Rp) {6\%};
    \node[font=\scriptsize, text=black!55] at (301.21:\Rp) {21\%};
    \node[font=\tiny, text=white] at (349.10:\Rp) {5\%};

    % Outer ring leader lines
    \def\Rl{2.15}
    \draw[thin, black!30] (13.34:\Ro) -- (2.20, 0.52);
    \node[font=\tiny, anchor=west, text=black!80] at (2.23, 0.52) {Parallel-3};
    \draw[thin, black!30] (100.61:\Ro) -- (100.61:\Rl);
    \node[font=\tiny, anchor=south east, text=black!80] at (105:\Rl) {Parallel-4};
    \draw[thin, black!30] (208.49:\Ro) -- (208.49:\Rl);
    \node[font=\tiny, anchor=east, text=black!80] at (208.49:\Rl) {HB-3x1};
    \draw[thin, black!30] (252.73:\Ro) -- (252.73:\Rl);
    \node[font=\tiny, anchor=north east, text=black!80] at (252.73:\Rl) {HB-3x2+};
    \draw[thin, black!30] (301.21:\Ro) -- (301.21:\Rl);
    \node[font=\tiny, anchor=north west, text=black!80] at (301.21:\Rl) {HB-4x1};
    \draw[thin, black!30] (340:\Ro) -- (2.20, -0.55);
    \node[font=\tiny, anchor=west, text=black!60] at (2.23, -0.55) {HB-4x2+};
    \draw[thin, black!30] (349.10:\Ro) -- (2.20, -0.22);
    \node[font=\tiny, anchor=west, text=black!80] at (2.23, -0.22) {Serial-2x1};
    \draw[thin, black!30] (358.79:\Ro) -- (2.20, 0.12);
    \node[font=\tiny, anchor=west, text=black!60] at (2.23, 0.12) {Serial-2x2+};
    \end{scope}

    % ============ Right-side five-column table ============
    \begin{scope}[shift={(1.5, 1.2)}]
      \def\colA{0}
      \def\colB{0.30}
      \def\colC{1.80}
      \def\colD{3.50}
      \def\colE{4.30}

      % --- Header ---
      \node[anchor=west, font=\tiny\bfseries, text=black!65] at (\colB, 0) {Type};
      \node[anchor=west, font=\tiny\bfseries, text=black!65] at (\colC, 0) {Subtype};
      \node[anchor=east, font=\tiny\bfseries, text=black!65] at (\colD, 0) {Count};
      \node[anchor=east, font=\tiny\bfseries, text=black!65] at (\colE, 0) {Total};
      \draw[black!25, thin] (\colA, -0.12) -- (\colE, -0.12);

      % --- Parallel-only: 2 subtypes ---
      \def\rh{0.28}
      \fill[cPar] (\colA, {-0.25 - 0.5*\rh - 0.08}) rectangle ++(0.18, 0.14);
      \node[anchor=west, font=\tiny, text=black!90] at (\colB, {-0.25 - 0.5*\rh}) {Parallel-only};
      \node[anchor=west, font=\tiny, text=black!80] at (\colC, -0.25) {Parallel-3};
      \node[anchor=east, font=\tiny, text=black!80] at (\colD, -0.25) {22};
      \node[anchor=west, font=\tiny, text=black!80] at (\colC, {-0.25 - \rh}) {Parallel-4};
      \node[anchor=east, font=\tiny, text=black!80] at (\colD, {-0.25 - \rh}) {122};
      \node[anchor=east, font=\tiny\bfseries, text=black!90] at (\colE, {-0.25 - 0.5*\rh}) {144};

      % --- Separator ---
      \pgfmathsetmacro{\sepA}{-0.25 - 2*\rh + 0.06}
      \draw[black!15, thin] (\colA, \sepA) -- (\colE, \sepA);

      % --- Hybrid-balanced: 4 subtypes ---
      \pgfmathsetmacro{\hStart}{-0.25 - 2*\rh - 0.06}
      \fill[cHyb] (\colA, {\hStart - 1.5*\rh - 0.08}) rectangle ++(0.18, 0.14);
      \node[anchor=west, font=\tiny, text=black!90] at (\colB, {\hStart - 1.5*\rh}) {Hybrid-balanced};
      \node[anchor=west, font=\tiny, text=black!80] at (\colC, \hStart) {HB-3x1};
      \node[anchor=east, font=\tiny, text=black!80] at (\colD, \hStart) {56};
      \node[anchor=west, font=\tiny, text=black!80] at (\colC, {\hStart - \rh}) {HB-3x2+};
      \node[anchor=east, font=\tiny, text=black!80] at (\colD, {\hStart - \rh}) {17};
      \node[anchor=west, font=\tiny, text=black!80] at (\colC, {\hStart - 2*\rh}) {HB-4x1};
      \node[anchor=east, font=\tiny, text=black!80] at (\colD, {\hStart - 2*\rh}) {63};
      \node[anchor=west, font=\tiny, text=black!80] at (\colC, {\hStart - 3*\rh}) {HB-4x2+};
      \node[anchor=east, font=\tiny, text=black!80] at (\colD, {\hStart - 3*\rh}) {1};
      \node[anchor=east, font=\tiny\bfseries, text=black!90] at (\colE, {\hStart - 1.5*\rh}) {137};

      % --- Separator ---
      \pgfmathsetmacro{\sepB}{\hStart - 4*\rh + 0.06}
      \draw[black!15, thin] (\colA, \sepB) -- (\colE, \sepB);

      % --- Serial-dominant: 2 subtypes ---
      \pgfmathsetmacro{\sStart}{\hStart - 4*\rh - 0.06}
      \fill[cSer] (\colA, {\sStart - 0.5*\rh - 0.08}) rectangle ++(0.18, 0.14);
      \node[anchor=west, font=\tiny, text=black!90] at (\colB, {\sStart - 0.5*\rh}) {Serial-dominant};
      \node[anchor=west, font=\tiny, text=black!80] at (\colC, \sStart) {Serial-2x1};
      \node[anchor=east, font=\tiny, text=black!80] at (\colD, \sStart) {14};
      \node[anchor=west, font=\tiny, text=black!80] at (\colC, {\sStart - \rh}) {Serial-2x2+};
      \node[anchor=east, font=\tiny, text=black!80] at (\colD, {\sStart - \rh}) {2};
      \node[anchor=east, font=\tiny\bfseries, text=black!90] at (\colE, {\sStart - 0.5*\rh}) {16};
    \end{scope}

  \end{tikzpicture}
  \caption{Two-layer distribution of reasoning types and subtypes in PJB queries.}
  \label{fig:reasoning-type-pie-fixed}
\end{figure}

\begin{sloppypar}
Below the three reasoning types, PJB further derives finer reasoning subtypes from the numerical combination of parallel width $\times$ serial depth. Figure~\ref{fig:reasoning-type-pie-fixed} displays the complete hierarchical distribution as a two-layer ring chart: the inner ring shows the three major reasoning types, and the outer ring expands into eight reasoning subtypes---for example, Parallel-3 (three-way parallel, no serial), HB-4x1 (four-way parallel + one serial step), and Serial-2x2+ (two-way parallel + two or more serial steps). It should be noted that all reasoning labels are rule-based heuristic estimates used for diagnosing system performance by task complexity; they are not manually annotated reasoning traces, nor are they used as supervision targets.
\end{sloppypar}

\subsection{Compliance, Fairness, and Data Privacy}
\label{sec:compliance}

\begin{figure}[ht]
  \centering
  \resizebox{\columnwidth}{!}{%
  \begin{tikzpicture}[
    node distance=0.45cm and 0.55cm,
    base/.style={
        draw, thick, rounded corners=2pt,
        align=center, font=\rmfamily\scriptsize,
        inner sep=2.5pt, minimum height=0.7cm,
        fill=white
    },
    phaseNode/.style={base, minimum width=1.9cm, fill=paperBlue!6, draw=paperBlue!80},
    guardNode/.style={base, minimum width=1.9cm, fill=paperOrange!8, draw=paperOrange!80},
    outNode/.style={base, minimum width=1.9cm, fill=paperGreen!8, draw=paperGreen!80, line width=1pt},
    arr/.style={-{Stealth[scale=0.8]}, thick, draw=black!60},
    lbl/.style={font=\tiny, text=black!55, align=center}
  ]
    % Upper row: data flow
    \node[phaseNode] (src) {\textbf{Raw Data}};
    \node[phaseNode, right=of src] (deident) {\textbf{De-identification}};
    \node[phaseNode, right=of deident] (reldef) {\textbf{Relevance Def.}};
    \node[phaseNode, right=of reldef] (label) {\textbf{Label Design}};
    \node[outNode, right=of label] (release) {\textbf{Internal Release}};

    \draw[arr] (src) -- (deident);
    \draw[arr] (deident) -- (reldef);
    \draw[arr] (reldef) -- (label);
    \draw[arr] (label) -- (release);

    % Lower row: safeguards
    \node[guardNode, below=of deident] (g1) {Remove PII\\[-1pt]\tiny Data Minimization};
    \node[guardNode, below=of reldef] (g2) {Anchor on Competency\\[-1pt]\tiny Exclude Sensitive Attr.};
    \node[guardNode, below=of label] (g3) {Domain/Reasoning\\[-1pt]\tiny Indep. of Sensitive Attr.};
    \node[guardNode, below=of release] (g4) {Usage Restriction\\[-1pt]\tiny Internal Governance};

    \draw[arr, dashed, draw=black!40] (deident) -- (g1);
    \draw[arr, dashed, draw=black!40] (reldef) -- (g2);
    \draw[arr, dashed, draw=black!40] (label) -- (g3);
    \draw[arr, dashed, draw=black!40] (release) -- (g4);

    % Bottom labels
    \node[lbl, below=0.12cm of g1] {Privacy};
    \node[lbl, below=0.12cm of g2] {Fairness};
    \node[lbl, below=0.12cm of g3] {Fairness};
    \node[lbl, below=0.12cm of g4] {Compliance};
  \end{tikzpicture}%
  }
  \caption{Compliance, fairness, and data privacy safeguards in the PJB construction pipeline.}
  \label{fig:compliance-flow}
\end{figure}
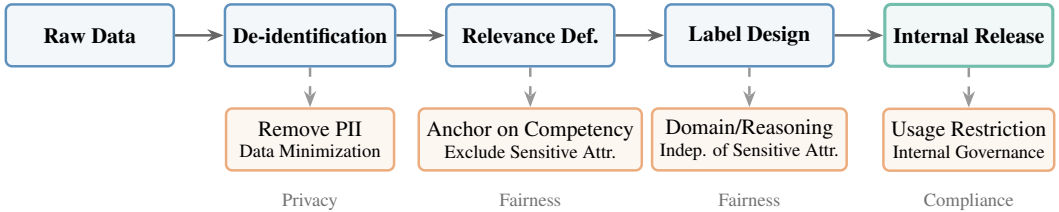

Constructing a retrieval benchmark from real job seekers and real job postings means that privacy, fairness, and compliance safeguards must be embedded at every stage of the data lifecycle rather than applied as an afterthought. PJB establishes four lines of defense along the construction pipeline (Figure~\ref{fig:compliance-flow}), ensuring that safeguards advance in lockstep with the data flow.

The first line of defense is at the data ingestion point: all JDs and CVs undergo de-identification before entering the benchmark, removing names, contact information, and other personally identifiable information, retaining only competency and experience fields needed for evaluation. This practice follows the principle of data minimization and is consistent with anonymized resume dataset practices designed for career trajectory modeling \citep{yamashita2024openresume}.

De-identification addresses the ``whose data'' question, but recruitment scenarios also face fairness risks in ``how to judge''---existing research has shown that retrieval-based or LLM-based resume screening systems may exhibit statistical disparities across gender, race, and other dimensions \citep{wilson2024gender,vladimirova2024fairjob}. Accordingly, PJB anchors its relevance definition on job competency rather than education level or demographic attributes at the relevance definition stage, limiting bias from entering the label system at the source. This design choice naturally extends to the diagnostic labels: domain families aggregate by core matching competency, and reasoning types partition by query structural complexity, both independent of candidate sensitive attributes, ensuring that diagnostic analysis provides sliced insights without introducing new fairness risks.

Finally, PJB explicitly restricts usage at the release boundary: the benchmark is designated for offline evaluation and internal research only, with data sources and usage scope governed by internal policies and applicable regulations. It should be emphasized that PJB, as a diagnostic tool for retrieval systems, does not replace or diminish the human oversight and compliance audit responsibilities that employers must bear when deploying automated recruitment tools.

\subsection{Evaluation Protocol}

PJB's primary evaluation task is full-corpus retrieval: the system produces a scored ranking list for each JD over the fixed resume corpus. This protocol supports sparse retrieval, dense retrieval, late-interaction retrieval, and hybrid retrieval. Additionally, PJB allows reranking results on a fixed candidate set as a separate diagnostic track, but such results should only be directly compared when candidate sources and candidate depths are consistent; they should not be unconditionally mixed with full-corpus recall results in a single overall leaderboard. Since the standardized runs used in this paper primarily have depth 20, we only report metrics at cutoff points not exceeding 20.

For metrics, PJB uses nDCG@10 as the official primary metric, both because nDCG is a standard gain-based ranking metric \citep{jarvelin2002cumulated} and because modern retrieval benchmarks commonly adopt nDCG@10 as their primary score \citep{muennighoff-etal-2023-mteb,su2024bright}. To supplement early-rank coverage and first-hit position, we also report Recall@20 and MRR@10. All metrics are computed per query and then macro-averaged across queries; this avoids high-positive queries dominating the overall score and is more suitable for multi-positive person-job retrieval scenarios. Since the released relevance judgments contain only positives, all missing pairs are treated as zero gain at evaluation time, consistent with TREC-style ranking evaluation practice \citep{voorhees2005trec}.

Beyond the overall primary metric, PJB's diagnostic analysis includes three types of supplementary reports. The first is grouped evaluation by domain family, reasoning type, and reasoning subtype, used to identify on which business slices a system consistently benefits or consistently fails. The second is auxiliary metrics at the top-20 view, such as nDCG@20, Precision@20, and HitRate@20. The third is low-gain query profiling, which classifies queries into zero-gain, low-gain, and other types to observe whether a system persistently ``completely fails'' or ``yields only marginal benefit'' on a subset of queries. Thus, PJB's methodological focus is not merely to define a retrieval score, but to define a unified evaluation protocol that simultaneously supports aggregate comparison, stratified analysis, and error profiling.

\section{Results}
The dense retrieval experiments completed to date employ two baseline models---the in-house CRE-T1-0.6B and the general-purpose Qwen3-Embedding-0.6B---each augmented with query understanding (QU) and reranking (Rerank) modules, forming a 2$\times$4 ablation matrix of 8 runs. Results show that the reranking module yields stable positive gains only on CRE-T1, while degrading performance on Qwen3 across all combinations; performance heterogeneity across domain families and reasoning types remains pronounced, so results must be interpreted through a sliced perspective. Unless otherwise stated, all conclusions in this section are restricted to the current dense retrieval runs and do not extrapolate to BM25 recall pipelines, hybrid pipelines, or more general end-to-end systems.

\subsection{Overall Comparison}

Table~\ref{tab:overall-six-runs} summarizes the overall results of two retrieval models (in-house CRE-T1-0.6B and general-purpose Qwen3-Embedding-0.6B) under different module combinations. The QU module uses Qwen3-8B as the query rewriting model, and the Rerank module uses Qwen3-Reranker-8B. The comparison between the two model families reveals three key phenomena: (1) CRE-T1-0.6B's baseline nDCG@10 (0.2070) is approximately 3.5$\times$ that of Qwen3-Embedding-0.6B (0.0592), indicating that domain-adapted training is critical for recruitment retrieval; (2) on CRE-T1-0.6B, the Rerank module provides stable positive gains (+8.9\%), whereas on Qwen3-Embedding-0.6B, general-purpose reranking actually degrades performance ($-$26.1\%); (3) the QU module yields negative gains for both models, and the combined QU + Rerank effect is worse than Rerank alone, suggesting that the current query rewriting strategy may lose the structural matching information inherent in original job descriptions.

\begin{table}[ht]
  \centering
  \caption{Overall results of in-house and general-purpose models under different module combinations. QU = Qwen3-8B query rewriting, Rerank = Qwen3-Reranker-8B. Primary metric is nDCG@10; bold indicates column best.}
  \label{tab:overall-six-runs}
  \small
  \begin{adjustbox}{width=\textwidth}
  \begin{tabular}{l c c c c}
    \toprule
    \textbf{Run} & \textbf{nDCG@10} & \textbf{nDCG@20} & \textbf{Recall@20} & \textbf{MRR@10} \\
    \midrule
    CRE-T1-0.6B & 0.2070 & 0.2303 & 0.3183 & 0.3204 \\
    CRE-T1-0.6B + QU & 0.1887 & 0.2007 & 0.2490 & 0.3059 \\
    CRE-T1-0.6B + Rerank & \textbf{0.2253} & \textbf{0.2555} & \textbf{0.3415} & \textbf{0.3612} \\
    CRE-T1-0.6B + QU + Rerank & 0.1423 & 0.1565 & 0.2031 & 0.2402 \\
    \midrule
    Qwen3-Embedding-0.6B & 0.0592 & 0.0779 & 0.1407 & 0.0840 \\
    Qwen3-Embedding-0.6B + QU & 0.0444 & 0.0649 & 0.1195 & 0.0673 \\
    Qwen3-Embedding-0.6B + Rerank & 0.0437 & 0.0546 & 0.0889 & 0.0783 \\
    Qwen3-Embedding-0.6B + QU + Rerank & 0.0336 & 0.0481 & 0.0850 & 0.0614 \\
    \bottomrule
  \end{tabular}
  \end{adjustbox}
\end{table}

Table~\ref{tab:dense-v1-ablation} further compares the module gain deltas ($\Delta$) for both model families. On CRE-T1-0.6B, Rerank is the only module that provides positive gains (nDCG@10 improvement of +0.0184), while QU and QU+Rerank both yield negative gains. On Qwen3-Embedding-0.6B, all module combinations produce negative gains, exhibiting compounding degradation: QU and Rerank each cause approximately $-$25\% decline, and their combination amplifies the decline to $-$43\%. This indicates that general-purpose enhancement modules not only fail to compensate for retriever weaknesses in the vertical recruitment domain but actually introduce additional noise.

\begin{table}[ht]
  \centering
  \caption{Module combination gains relative to respective baselines. Positive values indicate improvement over baseline; negative values indicate degradation.}
  \label{tab:dense-v1-ablation}
  \small
  \begin{adjustbox}{width=\textwidth}
  \begin{tabular}{l c c c c c c}
    \toprule
    \textbf{Run} & \textbf{nDCG@10} & \textbf{$\Delta$ nDCG@10} & \textbf{nDCG@20} & \textbf{$\Delta$ nDCG@20} & \textbf{Recall@20} & \textbf{$\Delta$ Recall@20} \\
    \midrule
    \multicolumn{7}{l}{\textit{CRE-T1-0.6B baseline: nDCG@10 = 0.2070}} \\
    \quad + QU & 0.1887 & $-$0.0183 & 0.2007 & $-$0.0295 & 0.2490 & $-$0.0692 \\
    \quad + Rerank & \textbf{0.2253} & \textbf{+0.0184} & \textbf{0.2555} & \textbf{+0.0252} & \textbf{0.3415} & \textbf{+0.0233} \\
    \quad + QU + Rerank & 0.1423 & $-$0.0646 & 0.1565 & $-$0.0738 & 0.2031 & $-$0.1152 \\
    \midrule
    \multicolumn{7}{l}{\textit{Qwen3-Embedding-0.6B baseline: nDCG@10 = 0.0592}} \\
    \quad + QU & 0.0444 & $-$0.0148 & 0.0649 & $-$0.0130 & 0.1195 & $-$0.0212 \\
    \quad + Rerank & 0.0437 & $-$0.0155 & 0.0546 & $-$0.0234 & 0.0889 & $-$0.0518 \\
    \quad + QU + Rerank & 0.0336 & $-$0.0256 & 0.0481 & $-$0.0298 & 0.0850 & $-$0.0557 \\
    \bottomrule
  \end{tabular}
  \end{adjustbox}
\end{table}

\subsection{Domain-Family Heterogeneity}

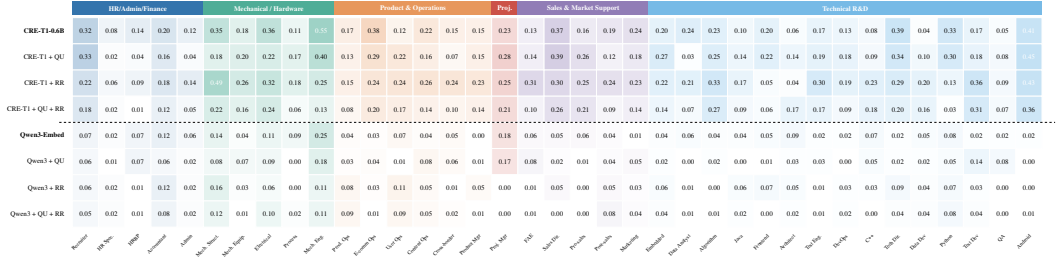
\begin{figure}[ht]
  \centering
  \begin{adjustbox}{width=\textwidth}
  \begin{tikzpicture}[x=1.2cm, y=1.2cm]
    % --- Colors consistent with domain families ---
    \definecolor{colorHR}{RGB}{52, 116, 172}
    \definecolor{colorMech}{RGB}{79, 173, 151}
    \definecolor{colorProd}{RGB}{231, 151, 92}
    \definecolor{colorProj}{RGB}{204, 82, 72}
    \definecolor{colorSales}{RGB}{145, 125, 182}
    \definecolor{colorTech}{RGB}{119, 186, 230}

    \newcommand{\heatmapcell}[4]{%
      \pgfmathtruncatemacro{\intensity}{#4*100}%
      \fill[#3!\intensity!white, draw=white, line width=0.5pt] (#1,#2) rectangle (#1+1,#2+1);
      \pgfmathtruncatemacro{\useblack}{#4 > 0.4 ? 0 : 1}%
      \node[font=\footnotesize] at ({#1+0.5},{#2+0.5}) {\ifnum\useblack=1 \color{black}\else \color{white}\fi #4};
    }

    % --- Top: 6 domain family indicator bars ---
    \begin{scope}[every node/.style={font=\small\bfseries, text=white, inner sep=2pt}]
      \fill[colorHR] (0,8.1) rectangle (5,8.6) node[midway] {HR/Admin/Finance};
      \fill[colorMech] (5,8.1) rectangle (10,8.6) node[midway] {Mechanical / Hardware};
      \fill[colorProd] (10,8.1) rectangle (16,8.6) node[midway] {Product \& Operations};
      \fill[colorProj] (16,8.1) rectangle (17,8.6) node[midway] {Proj.};
      \fill[colorSales] (17,8.1) rectangle (22,8.6) node[midway] {Sales \& Market Support};
      \fill[colorTech] (22,8.1) rectangle (37,8.6) node[midway] {Technical R\&D};
    \end{scope}

    % --- Data: columns=37 job categories, rows=8 runs ---
    % Row order top to bottom: CRE-T1, +QU, +RR, +QU+RR, Qwen3, +QU, +RR, +QU+RR
    % HR/Admin/Finance (0--4)
    \foreach \x/\va/\vb/\vc/\vd/\ve/\vf/\vg/\vh in {
      0/0.32/0.33/0.22/0.18/0.07/0.06/0.06/0.05,
      1/0.08/0.02/0.06/0.02/0.02/0.01/0.02/0.02,
      2/0.14/0.04/0.09/0.01/0.07/0.07/0.01/0.01,
      3/0.20/0.16/0.18/0.12/0.12/0.06/0.12/0.08,
      4/0.12/0.04/0.14/0.05/0.06/0.02/0.02/0.02
    } {
      \heatmapcell{\x}{7}{colorHR}{\va}\heatmapcell{\x}{6}{colorHR}{\vb}\heatmapcell{\x}{5}{colorHR}{\vc}\heatmapcell{\x}{4}{colorHR}{\vd}\heatmapcell{\x}{3}{colorHR}{\ve}\heatmapcell{\x}{2}{colorHR}{\vf}\heatmapcell{\x}{1}{colorHR}{\vg}\heatmapcell{\x}{0}{colorHR}{\vh}
    }
    % Mechanical / Hardware (5--9)
    \foreach \x/\va/\vb/\vc/\vd/\ve/\vf/\vg/\vh in {
      5/0.35/0.18/0.49/0.22/0.14/0.08/0.16/0.12,
      6/0.18/0.20/0.26/0.16/0.04/0.07/0.03/0.01,
      7/0.36/0.22/0.32/0.24/0.11/0.09/0.06/0.10,
      8/0.11/0.17/0.18/0.06/0.09/0.00/0.00/0.02,
      9/0.55/0.40/0.25/0.13/0.25/0.18/0.11/0.11
    } {
      \heatmapcell{\x}{7}{colorMech}{\va}\heatmapcell{\x}{6}{colorMech}{\vb}\heatmapcell{\x}{5}{colorMech}{\vc}\heatmapcell{\x}{4}{colorMech}{\vd}\heatmapcell{\x}{3}{colorMech}{\ve}\heatmapcell{\x}{2}{colorMech}{\vf}\heatmapcell{\x}{1}{colorMech}{\vg}\heatmapcell{\x}{0}{colorMech}{\vh}
    }
    % Product & Operations (10--15)
    \foreach \x/\va/\vb/\vc/\vd/\ve/\vf/\vg/\vh in {
      10/0.17/0.13/0.15/0.08/0.04/0.03/0.08/0.09,
      11/0.38/0.29/0.24/0.20/0.03/0.04/0.03/0.01,
      12/0.12/0.22/0.24/0.17/0.07/0.01/0.11/0.09,
      13/0.22/0.16/0.26/0.14/0.04/0.08/0.05/0.05,
      14/0.15/0.07/0.24/0.10/0.05/0.06/0.01/0.02,
      15/0.15/0.15/0.23/0.14/0.00/0.01/0.05/0.01
    } {
      \heatmapcell{\x}{7}{colorProd}{\va}\heatmapcell{\x}{6}{colorProd}{\vb}\heatmapcell{\x}{5}{colorProd}{\vc}\heatmapcell{\x}{4}{colorProd}{\vd}\heatmapcell{\x}{3}{colorProd}{\ve}\heatmapcell{\x}{2}{colorProd}{\vf}\heatmapcell{\x}{1}{colorProd}{\vg}\heatmapcell{\x}{0}{colorProd}{\vh}
    }
    % Project Management (16)
    \heatmapcell{16}{7}{colorProj}{0.23}\heatmapcell{16}{6}{colorProj}{0.28}\heatmapcell{16}{5}{colorProj}{0.25}\heatmapcell{16}{4}{colorProj}{0.21}\heatmapcell{16}{3}{colorProj}{0.18}\heatmapcell{16}{2}{colorProj}{0.17}\heatmapcell{16}{1}{colorProj}{0.00}\heatmapcell{16}{0}{colorProj}{0.00}
    % Sales & Market Support (17--21)
    \foreach \x/\va/\vb/\vc/\vd/\ve/\vf/\vg/\vh in {
      17/0.13/0.14/0.31/0.10/0.06/0.08/0.01/0.01,
      18/0.37/0.39/0.30/0.26/0.05/0.02/0.05/0.00,
      19/0.16/0.26/0.25/0.21/0.06/0.01/0.00/0.00,
      20/0.19/0.12/0.24/0.09/0.04/0.04/0.05/0.08,
      21/0.24/0.18/0.23/0.14/0.01/0.05/0.03/0.04
    } {
      \heatmapcell{\x}{7}{colorSales}{\va}\heatmapcell{\x}{6}{colorSales}{\vb}\heatmapcell{\x}{5}{colorSales}{\vc}\heatmapcell{\x}{4}{colorSales}{\vd}\heatmapcell{\x}{3}{colorSales}{\ve}\heatmapcell{\x}{2}{colorSales}{\vf}\heatmapcell{\x}{1}{colorSales}{\vg}\heatmapcell{\x}{0}{colorSales}{\vh}
    }
    % Technical R&D (22--36)
    \foreach \x/\va/\vb/\vc/\vd/\ve/\vf/\vg/\vh in {
      22/0.20/0.27/0.22/0.14/0.04/0.02/0.06/0.04,
      23/0.24/0.03/0.21/0.07/0.06/0.00/0.01/0.01,
      24/0.23/0.25/0.33/0.27/0.04/0.02/0.00/0.01,
      25/0.10/0.14/0.17/0.09/0.04/0.00/0.06/0.02,
      26/0.20/0.22/0.05/0.06/0.05/0.01/0.07/0.00,
      27/0.06/0.14/0.04/0.17/0.09/0.03/0.05/0.02,
      28/0.17/0.19/0.30/0.17/0.02/0.03/0.01/0.01,
      29/0.13/0.18/0.19/0.09/0.02/0.00/0.03/0.02,
      30/0.08/0.09/0.23/0.18/0.07/0.05/0.03/0.00,
      31/0.39/0.34/0.29/0.20/0.02/0.02/0.09/0.04,
      32/0.04/0.10/0.20/0.16/0.05/0.02/0.04/0.04,
      33/0.33/0.30/0.13/0.03/0.08/0.05/0.07/0.08,
      34/0.17/0.18/0.36/0.31/0.02/0.14/0.03/0.04,
      35/0.05/0.08/0.09/0.07/0.02/0.08/0.00/0.00,
      36/0.41/0.45/0.43/0.36/0.02/0.00/0.00/0.01
    } {
      \heatmapcell{\x}{7}{colorTech}{\va}\heatmapcell{\x}{6}{colorTech}{\vb}\heatmapcell{\x}{5}{colorTech}{\vc}\heatmapcell{\x}{4}{colorTech}{\vd}\heatmapcell{\x}{3}{colorTech}{\ve}\heatmapcell{\x}{2}{colorTech}{\vf}\heatmapcell{\x}{1}{colorTech}{\vg}\heatmapcell{\x}{0}{colorTech}{\vh}
    }

    % --- Y-axis: eight runs ---
    \node[anchor=east, font=\small\bfseries] at (-0.2, 7.5) {CRE-T1-0.6B};
    \node[anchor=east, font=\small] at (-0.2, 6.5) {CRE-T1 + QU};
    \node[anchor=east, font=\small] at (-0.2, 5.5) {CRE-T1 + RR};
    \node[anchor=east, font=\small] at (-0.2, 4.5) {CRE-T1 + QU + RR};
    \node[anchor=east, font=\small\bfseries] at (-0.2, 3.5) {Qwen3-Embed};
    \node[anchor=east, font=\small] at (-0.2, 2.5) {Qwen3 + QU};
    \node[anchor=east, font=\small] at (-0.2, 1.5) {Qwen3 + RR};
    \node[anchor=east, font=\small] at (-0.2, 0.5) {Qwen3 + QU + RR};

    % --- Separator ---
    \draw[black, thick, dashed] (-0.5, 4) -- (37.5, 4);

    % --- X-axis: job categories ---
    \foreach \x/\txt in {
      0.5/Recruiter, 1.5/HR Spec., 2.5/HRBP, 3.5/Accountant, 4.5/Admin,
      5.5/Mech. Struct., 6.5/Mech. Equip., 7.5/Electrical, 8.5/Process, 9.5/Mech. Eng.,
      10.5/Prod. Ops, 11.5/E-comm Ops, 12.5/User Ops, 13.5/Content Ops, 14.5/Cross-border, 15.5/Product Mgr,
      16.5/Proj. Mgr,
      17.5/FAE, 18.5/Sales Dir., 19.5/Pre-sales, 20.5/Post-sales, 21.5/Marketing,
      22.5/Embedded, 23.5/Data Analyst, 24.5/Algorithm, 25.5/Java, 26.5/Frontend, 27.5/Architect,
      28.5/Test Eng., 29.5/DevOps, 30.5/C++, 31.5/Tech Dir., 32.5/Data Dev, 33.5/Python,
      34.5/Test Dev, 35.5/QA, 36.5/Android
    } {
      \node[anchor=north east, rotate=45, font=\scriptsize] at (\x, -0.1) {\txt};
    }
  \end{tikzpicture}
  \end{adjustbox}
  \caption{nDCG@10 heatmap of eight runs across six domain families and 37 job categories. The x-axis shows job categories grouped by domain family; the y-axis shows runs, with the dashed line separating the in-house CRE-T1-0.6B series (top) from the general-purpose Qwen3-Embedding-0.6B series (bottom). RR = Qwen3-Reranker-8B, QU = Qwen3-8B query rewriting. Darker colors indicate higher scores.}
  \label{fig:domain-heatmap-six-runs}
\end{figure}

Beyond overall results, domain-family-level differences reveal deeper disparities between the in-house and general-purpose models. Figure~\ref{fig:domain-heatmap-six-runs} presents the nDCG@10 heatmap of all 8 runs across the six domain families and 37 job categories. CRE-T1-0.6B significantly outperforms Qwen3-Embedding-0.6B in nearly all job categories, with the latter's lower half of the heatmap almost entirely light-colored (nDCG@10 $<$ 0.10). In terms of module gains, CRE-T1-0.6B + Rerank achieves the largest improvement in the Sales \& Market Support domain (e.g., FAE from 0.13 to 0.31), while Qwen3-Embedding-0.6B + Rerank drops nDCG@10 to 0.00 in the Project Management category, suggesting that general-purpose reranking models may completely fail on low-sample categories.

\subsection{Reasoning-Type Heterogeneity}

Reasoning types further reveal structural differences in module gains (Table~\ref{tab:reasoning-six-runs}). For CRE-T1-0.6B, the gains from reranking alone concentrate on serial-dominant queries: nDCG@10 rises from the baseline 0.1879 to 0.3809 ($+$0.1930), far exceeding the $+$0.0313 gain on hybrid-balanced queries and actually declining on parallel-only queries ($-$0.0133). This suggests that the reranking module primarily addresses the ranking problem of ``needing more serial inference to promote positives to the top.'' Adding the query understanding module alone shows similarly uneven behavior: only serial-dominant queries receive positive gains ($+$0.0825), while hybrid-balanced and parallel-only types both decline. The QU + Rerank combination fails to preserve the rerank-only advantage on serial-dominant queries, indicating that the two modules do not form a stable combined effect under the current setting.

The Qwen3-Embedding-0.6B series achieves significantly lower absolute scores than CRE-T1 across all reasoning types, and both QU and Rerank modules produce universally negative effects for Qwen3: the best baseline on serial-dominant queries (0.0710) drops to 0.0053 after stacking QU + Rerank, nearly reaching zero. This further confirms that module gains are highly dependent on the base retriever's quality---when baseline recall is insufficient, post-processing modules cannot compensate and may even introduce additional noise.

\begin{table}[ht]
  \centering
  \caption{nDCG@10 of eight runs across three reasoning types. RR = Qwen3-Reranker-8B, QU = Qwen3-8B query rewriting.}
  \label{tab:reasoning-six-runs}
  \small
  \begin{adjustbox}{width=\textwidth}
  \begin{tabular}{l c c c c c c c c}
    \toprule
    & \multicolumn{4}{c}{\textbf{CRE-T1-0.6B}} & \multicolumn{4}{c}{\textbf{Qwen3-Embedding-0.6B}} \\
    \cmidrule(lr){2-5} \cmidrule(lr){6-9}
    \textbf{Reasoning Type} & \textbf{Base} & \textbf{+QU} & \textbf{+RR} & \textbf{+QU+RR} & \textbf{Base} & \textbf{+QU} & \textbf{+RR} & \textbf{+QU+RR} \\
    \midrule
    Hybrid-balanced  & 0.1755 & 0.1537 & 0.2068 & 0.1368 & 0.0514 & 0.0378 & 0.0490 & 0.0404 \\
    parallel-only    & \textbf{0.2389} & 0.2128 & 0.2257 & 0.1385 & 0.0653 & 0.0498 & 0.0397 & 0.0303 \\
    serial-dominant  & 0.1879 & 0.2704 & \textbf{0.3809} & 0.2238 & 0.0710 & 0.0512 & 0.0344 & 0.0053 \\
    \bottomrule
  \end{tabular}
  \end{adjustbox}
\end{table}

\subsection{Error Profile}

Error analysis yields conclusions consistent with the primary metric analysis (Table~\ref{tab:error-six-runs}). Among the CRE-T1-0.6B series, reranking alone reduces the bad query rate from the baseline 0.4141 to 0.3502 and the zero-gain rate from 0.2997 to 0.2391, making it the only configuration among the four CRE-T1 variants that positively improves the error profile. In contrast, QU and QU + Rerank combinations both increase the bad query rate, with the latter reaching 0.5354. The Qwen3-Embedding-0.6B series has a substantially higher overall bad query rate than CRE-T1 (baseline already at 0.7239), and QU and Rerank modules likewise fail to reduce it; QU + Rerank stacking further worsens it to 0.8182. This again demonstrates that module gains are highly dependent on base retriever quality: when baseline recall is inadequate, post-processing modules not only fail to compensate but introduce more zero-gain queries.

\begin{table}[ht]
  \centering
  \caption{Low-gain error profile of eight runs. Bad query rate = zero-gain rate + low-gain rate, based on nDCG@20 $\leq$ 0.10 threshold.}
  \label{tab:error-six-runs}
  \small
  \begin{tabular}{l c c c}
    \toprule
    \textbf{Run} & \textbf{bad\_query\_rate} & \textbf{ZeroGain\_rate} & \textbf{LowGain\_rate} \\
    \midrule
    CRE-T1-0.6B          & 0.4141 & 0.2997 & 0.1145 \\
    CRE-T1 + QU          & 0.4613 & 0.3906 & 0.0707 \\
    CRE-T1 + RR          & \textbf{0.3502} & \textbf{0.2391} & 0.1111 \\
    CRE-T1 + QU + RR     & 0.5354 & 0.4377 & 0.0976 \\
    \midrule
    Qwen3-Embed-0.6B     & 0.7239 & 0.5488 & 0.1751 \\
    Qwen3 + QU           & 0.7273 & 0.5926 & 0.1347 \\
    Qwen3 + RR           & 0.7980 & 0.6397 & 0.1582 \\
    Qwen3 + QU + RR      & 0.8182 & 0.6633 & 0.1549 \\
    \bottomrule
  \end{tabular}
\end{table}

Taken together, PJB not only distinguishes the overall performance differences between in-house and general-purpose retrieval models, but also reveals structural differences in module combinations across domain families, reasoning types, and bad-query profiles. The strongest conclusions supported by current evidence are: (1) CRE-T1-0.6B significantly outperforms Qwen3-Embedding-0.6B across all dimensions, demonstrating the necessity of domain-adapted training; (2) the reranking module is the only module that stably yields gains on CRE-T1; (3) QU and Rerank both produce negative effects on Qwen3, indicating that module gains are highly dependent on base retriever quality. These conclusions provide clear experimental priorities for subsequent system optimization.

\section{Discussion}
We argue that PJB's primary value lies not in providing yet another average-score leaderboard, but in transforming person-job retrieval into a structurally diagnosable evaluation plane. Compared with general-purpose benchmarks such as BEIR and MTEB that emphasize cross-dataset generalization and broad task coverage \citep{thakur2021beir,muennighoff-etal-2023-mteb}, and with reasoning-intensive benchmarks such as BRIGHT \citep{su2024bright}, PJB further anchors evaluation in real recruitment scenarios involving long documents, multiple constraints, and job-competency judgments. Current experiments, through the 2$\times$4 ablation matrix of in-house CRE-T1-0.6B and general-purpose Qwen3-Embedding-0.6B, reveal three key findings: aggregate metrics, domain slices, reasoning slices, and low-gain query profiles are complementary rather than substitutive---a single aggregate metric cannot fully expose a system's true weaknesses on business-critical queries; module gains are not inherently monotonic and are highly dependent on base retriever quality---the reranking module yields stable positive gains only on CRE-T1 while actually degrading performance on Qwen3; and domain-adapted training is essential for recruitment scenarios, with CRE-T1 substantially outperforming the equally-sized general-purpose model across all dimensions. The direct implication for system design is that the optimization priority for recruitment retrieval systems should not be placed solely on ``adding more upstream understanding modules'' but should first verify whether the base retriever is sufficiently strong, and only then validate which components genuinely improve ranking quality. Compared with multi-task HR data resources such as RJDB \citep{pezeshkpour2023rjdb}, PJB's most prominent utility lies in its ability to distinguish capability differences between models and module combinations in person-job matching scenarios through a unified retrieval protocol and diagnostic slicing.

At the same time, the extrapolation boundaries of the current discussion must be made explicit. First, the experiments in this paper cover only dense retrieval and therefore cannot be directly generalized to BM25 or hybrid pipelines. Second, the module combination evidence comes from two specific models (CRE-T1-0.6B and Qwen3-Embedding-0.6B) and cannot yet be generalized to all retrieval models. Third, some fine-grained domain buckets and reasoning buckets have limited sample sizes, making the observed phenomena more appropriately interpreted as empirical observations under controlled experiments rather than universal laws. Finally, while PJB's current relevance judgments and diagnostic labels are already sufficient to support diagnostic analysis, they remain constrained by positive-only sparsity, heuristic labeling, and internal data boundaries. In other words, the current results sufficiently demonstrate that PJB can expose system capability structures, but they are not yet sufficient to support strong extrapolation to broader recall pipelines or more complete system combinations; these limitations do not diminish PJB's value as a recruitment retrieval benchmark, but they clearly delimit the valid scope of current conclusions and provide clear direction for subsequent experimental priorities.

\section{Conclusion}
In summary, PJB v1.0 formalizes the matching of complete JDs and complete CVs as a reproducible, diagnostic offline retrieval evaluation dataset, and on top of the classical Cranfield/TREC evaluation paradigm, further advances the benchmark toward real recruitment scenarios involving long documents, multiple constraints, and job-competency judgments. In methodological positioning, it is adjacent to the retrieval benchmark lineage represented by BEIR, MTEB, and BRIGHT \citep{thakur2021beir,muennighoff-etal-2023-mteb,su2024bright}, but through domain-family and reasoning-type slicing, upgrades ``reporting aggregate scores'' to ``explaining system capability structures.'' On compliance, bias, and data privacy, PJB adopts de-identification and usage restriction during construction, defines relevance through job competency, and employs diagnostic labels independent of sensitive attributes, consistent with fairness and data minimization practices in recruitment scenarios. Current 2$\times$4 ablation experiments using in-house CRE-T1-0.6B and general-purpose Qwen3-Embedding-0.6B with QU and Rerank modules demonstrate that: person-job retrieval exhibits pronounced domain heterogeneity, reasoning heterogeneity, and low-gain query risks, making a single aggregate metric insufficient for characterizing system performance on real business queries; domain-adapted training is critical, with CRE-T1 substantially leading the general-purpose model across all dimensions; module gains are highly dependent on base retriever quality, with the reranking module stably yielding positive gains only on CRE-T1 while degrading Qwen3. Meanwhile, current conclusions remain primarily restricted to dense retrieval and two specific model combinations, and cannot be extrapolated to BM25, hybrid pipelines, or more complete end-to-end system matrices. Future work will continue to augment additional recall pipelines and system combinations, advance toward more natural-language query formulations, and further enhance the reproducibility documentation of benchmark construction details and annotation procedures.

\bibliographystyle{plainnat}

\appendix
\section{PJB v1.0 Example Data and Label Determination}
\label{app:examples}

This appendix illustrates the determination process for domain-family labels (\texttt{domain\_family}) and reasoning-type labels (\texttt{reasoning\_type}) in PJB v1.0 through three real query examples. It should be emphasized that both label types are \textbf{analysis labels} used for stratified diagnosis of evaluation results, not supervision signals for relevance judgment.

\subsection{Data Format Overview}

Each query is stored in JSONL format with the following top-level fields:

\begin{promptblock}
\begin{lstlisting}
{
  "query_id": 30,
  "text": "{ ... }",          // nested full job description
  "lang": "zh",
  "query_type": "jd2cv",
  "domain_family": "Mechanical / Hardware",
  "reasoning_profile": {
    "reasoning_type": "parallel-only",
    "parallel_width": 4,
    "serial_depth": 0,
    "reasoning_subtype": "Parallel-4"
  }
}
\end{lstlisting}
\end{promptblock}

The \texttt{text} field contains the complete job description structure, while \texttt{domain\_family} and \texttt{reasoning\_profile} are analysis labels automatically derived by rules. Relevance judgments are stored in a separate \texttt{qrels.tsv} file (binary annotation, 1 = relevant).

\subsection{Domain-Family Label Determination}

Domain-family labels are derived through \textbf{deterministic rule mapping} from the \texttt{work\_category} (job category) field, defining 6 domain families across 37 job categories. Example mapping rules:

\begin{itemize}[nosep]
  \item \texttt{work\_category} = ``Mechanical Structure Engineer'' $\rightarrow$ \texttt{domain\_family} = ``Mechanical / Hardware''
  \item \texttt{work\_category} = ``Test Engineer'' $\rightarrow$ \texttt{domain\_family} = ``Technical R\&D''
  \item \texttt{work\_category} = ``Cross-border E-commerce Operations'' $\rightarrow$ \texttt{domain\_family} = ``Product \& Operations''
\end{itemize}

This mapping is a one-to-one deterministic rule that does not depend on model inference or human judgment.

\subsection{Reasoning-Type Label Determination}

Reasoning-type labels are automatically derived from query content through \textbf{heuristic rules}, involving two dimensions:

\textbf{Parallel width} (\texttt{parallel\_width}): Counts the number of \textbf{explicit filtering conditions} in the query. Each satisfied condition adds +1:
\begin{itemize}[nosep]
  \item \texttt{location} (work location) is present
  \item \texttt{education} is present and not ``unrestricted''
  \item \texttt{salary} (salary range) is present
  \item Standardized \texttt{work\_years} is present (excluding sentinel value ``99'')
\end{itemize}

\textbf{Serial depth} (\texttt{serial\_depth}): Counts the number of \textbf{semantic inference} signals. Each satisfied condition adds +1:
\begin{itemize}[nosep]
  \item Responsibility description length $\geq$ 700 characters, or contains $\geq$ 10 domain skill keywords
  \item \texttt{work\_years} contains ``99'' (non-standard experience requirement, needs semantic normalization)
  \item Responsibility description contains age constraints (e.g., ``age not exceeding 35'')
\end{itemize}

The final reasoning type is jointly determined:
\begin{itemize}[nosep]
  \item parallel\_width $\geq$ 3 and serial\_depth = 0 $\Rightarrow$ \textbf{parallel-only}
  \item serial\_depth $\geq$ 1 and parallel\_width $\geq$ 3 $\Rightarrow$ \textbf{Hybrid-balanced}
  \item Otherwise $\Rightarrow$ \textbf{serial-dominant}
\end{itemize}

\subsection{Example 1: parallel-only Type}

\begin{promptblock}
\begin{lstlisting}
query_id: 30
domain_family: Mechanical / Hardware
reasoning_type: parallel-only
parallel_width: 4, serial_depth: 0

--- text (Job Description) ---
company_name: Goertek Inc.
job_title: Structural Design Engineer
work_category: Mechanical Structure Engineer
location: Shenzhen
education: Master's
work_years: 5-10 years
salary_range: 35-40K

responsibilities:
Job Responsibilities:
1. Interpret product SPEC, analyze ID drawings, and
   assess feasibility
2. Create 2D/3D drawings, prepare BOM, and output
   quotation materials
3. Cross-department communication with system, optical,
   hardware, and process engineers for design refinement
...
Requirements:
1. Education: Master's or above, mechanical engineering
2. Experience: 5+ years in consumer electronics
3. Proficient in Creo, AutoCAD, and other 3D/2D software
\end{lstlisting}
\end{promptblock}

\textbf{Determination process}:
\begin{enumerate}[nosep]
  \item \textbf{Domain family}: \texttt{work\_category} = ``Mechanical Structure Engineer'' $\in$ Mechanical / Hardware set, therefore \texttt{domain\_family} = ``Mechanical / Hardware.''
  \item \textbf{Parallel width}: location (Shenzhen) +1, education (Master's $\neq$ unrestricted) +1, salary (35--40K) +1, work\_years (5--10 years, no ``99'') +1 = \textbf{4}.
  \item \textbf{Serial depth}: Responsibility description length $<$ 700 characters, fewer than 10 skill keywords (Creo, AutoCAD, BOM, etc.), work\_years does not contain ``99,'' no age constraints = \textbf{0}.
  \item \textbf{Reasoning type}: parallel\_width = 4 $\geq$ 3 and serial\_depth = 0 $\Rightarrow$ \textbf{parallel-only} (Parallel-4).
\end{enumerate}

This query's matching primarily relies on parallel comparison of structured filtering conditions---location, education, salary, and experience---without requiring deep semantic inference.

\subsection{Example 2: Hybrid-balanced Type}

\begin{promptblock}
\begin{lstlisting}
query_id: 38
domain_family: Technical R&D
reasoning_type: Hybrid-balanced
parallel_width: 3, serial_depth: 1

--- text (Job Description) ---
company_name: Tesla (Shanghai) Co., Ltd.
job_title: Algorithm Test Engineer
work_category: Test Engineer
location: Shanghai
education: unrestricted
work_years: 3-5 years
salary_range: 18-22K

responsibilities:
Job Responsibilities:
1. Conduct autonomous driving system testing, including
   test case design, test plan design, and test system
   setup;
2. Build test datasets and construct automated testing
   frameworks;
3. Design and iterate test sets, and identify potential
   risks and bugs;
4. Locate issues and assist development teams in
   problem resolution;
5. Perform regression testing to ensure proper fixes;
6. Participate in requirement and technical reviews,
   and drive project quality improvement.

Requirements:
1. Bachelor's or above, CS or software major;
2. Proficient in test automation technologies, Linux
   environment design and development preferred;
3. Familiar with automated testing frameworks, Python
   and Java development experience preferred;
4. Autonomous driving system testing experience preferred.
\end{lstlisting}
\end{promptblock}

\textbf{Determination process}:
\begin{enumerate}[nosep]
  \item \textbf{Domain family}: \texttt{work\_category} = ``Test Engineer'' $\in$ Technical R\&D set.
  \item \textbf{Parallel width}: location (Shanghai) +1, education (unrestricted, \textbf{not counted}), salary (18--22K) +1, work\_years (3--5 years) +1 = \textbf{3}.
  \item \textbf{Serial depth}: Responsibility description contains numerous domain keywords ($\geq$ 10 matches: ``testing,'' ``algorithm,'' ``Python,'' ``Java,'' ``Linux,'' ``autonomous driving,'' ``data,'' etc.), triggering deep semantic signal +1 = \textbf{1}.
  \item \textbf{Reasoning type}: serial\_depth $\geq$ 1 and parallel\_width $\geq$ 3 $\Rightarrow$ \textbf{Hybrid-balanced} (HB-3x1).
\end{enumerate}

This query requires both parallel filtering on location, salary, and experience, and understanding the compound skill semantics of ``autonomous driving system testing + Python/Java + automation frameworks,'' making it a hybrid-balanced type.

\subsection{Example 3: serial-dominant Type}

\begin{promptblock}
\begin{lstlisting}
query_id: 1054
domain_family: Product & Operations
reasoning_type: serial-dominant
parallel_width: 2, serial_depth: 1

--- text (Job Description) ---
company_name: Shenzhen Leqi Network Technology Co., Ltd.
job_title: Overseas E-commerce Director (primarily Amazon,
           other channels secondary)
work_category: Cross-border E-commerce Operations
location: Shenzhen
education: unrestricted
work_years: 4-99 years
salary_range: 40-65K

responsibilities:
Job Responsibilities:
1. Oversee overseas e-commerce strategy, processes, and
   plans; coordinate cross-departmental resources to
   achieve self-operated e-commerce targets;
2. Develop promotional plans (new launches, major sales
   events, holiday promotions) and ensure revenue
   targets are met;
3. Analyze operations data across e-commerce platforms
   and adjust strategies accordingly; lead the team
   to achieve GMV targets;

Requirements:
1. Bachelor's or above, managed a team of at least 8
2. Industry unrestricted, but not considering apparel.
   Familiar with Amazon platform operations and market
   trends
3. Strong team management and training capabilities
\end{lstlisting}
\end{promptblock}

\textbf{Determination process}:
\begin{enumerate}[nosep]
  \item \textbf{Domain family}: \texttt{work\_category} = ``Cross-border E-commerce Operations'' $\in$ Product \& Operations set.
  \item \textbf{Parallel width}: location (Shenzhen) +1, education (unrestricted, not counted), salary (40--65K) +1, work\_years (4--99 years, contains sentinel ``99,'' \textbf{not counted} for parallel width) = \textbf{2}.
  \item \textbf{Serial depth}: work\_years contains ``99'' (non-standard experience requirement, requires semantic normalization to understand as ``4+ years'') +1 = \textbf{1}.
  \item \textbf{Reasoning type}: parallel\_width = 2 $<$ 3, does not satisfy parallel-only or Hybrid-balanced conditions $\Rightarrow$ \textbf{serial-dominant} (Serial-2x1).
\end{enumerate}

This query has few explicit filtering conditions (only location and salary), but the experience requirement needs semantic normalization (``4--99 years'' actually means ``4+ years with no upper limit''), and the responsibility description emphasizes strategic-level management competency assessment, making the matching more dependent on serial inference of deep job semantics.

\end{document}